\documentclass[aps,prl,superscriptaddress,twocolumn,nofootinbib,preprintnumbers]{revtex4}
\pdfoutput=1

\usepackage{color}
\usepackage{graphicx}
\usepackage[space]{grffile}

\usepackage{verbatim}
\usepackage{amsmath}
\usepackage{amssymb}
\usepackage{wasysym}
\usepackage[caption=false]{subfig}
\usepackage{url}
\usepackage{bbold}
\usepackage{slashed}
\usepackage{epstopdf}

\usepackage{multirow}
\usepackage{threeparttable}
\usepackage{paralist}

\usepackage{color}
\definecolor{darkgreen}{rgb}{0,0.5,0}
\definecolor{violet}{rgb}{0.5,0.5,1}

\newcommand{\FeH}{\text{[Fe/H]} }
\newcommand{\alphaFe}{\text{[$\alpha$/Fe]} }
\newcommand{\Eris}{\textsc{Eris} }

\DeclareRobustCommand{\Fig}[1]{Fig.~\ref{#1}}

\DeclareRobustCommand{\Ref}[1]{Ref.~\cite{#1}}

\newcommand{\be}{\begin{equation}}
\newcommand{\ee}{\end{equation}}
\newcommand{\bea}{\begin{eqnarray}}
\newcommand{\eea}{\end{eqnarray}}
\newcommand{\bi}{\begin{itemize}}
\newcommand{\ei}{\end{itemize}}

\usepackage{xspace}

\usepackage[colorlinks=true
,urlcolor=blue
,anchorcolor=blue
,citecolor=blue
,filecolor=blue
,linkcolor=blue
,menucolor=blue
,linktocpage=true
,pdfproducer=medialab
,pdfa=true
]{hyperref}

\begin{document}

\title{Empirical Determination of Dark Matter Velocities using Metal-Poor Stars}
\preprint{MIT-CTP/4899}

\author{Jonah Herzog-Arbeitman}
\email{jonahh@princeton.edu}
\affiliation{Department of Physics, Princeton University, Princeton, NJ 08544, USA}

\author{Mariangela Lisanti}
\email{mlisanti@princeton.edu}
\affiliation{Department of Physics, Princeton University, Princeton, NJ 08544, USA}

\author{Piero Madau}
\email{pmadau@ucolick.org}
\affiliation{Department of Astronomy \& Astrophysics, University of California, Santa Cruz, CA 95064, USA}
\affiliation{Institut d'Astrophysique de Paris, Sorbonne Universit{\'e}s, 75014 Paris, France}

\author{Lina Necib}
\email{lnecib@mit.edu}
\affiliation{Center for Theoretical Physics, Massachusetts Institute of Technology, Cambridge, MA 02139, USA}

\begin{abstract}

The Milky Way dark matter halo is formed from the accretion of smaller subhalos.  These sub-units also harbor stars---typically old and metal-poor---that are deposited in the Galactic inner regions by disruption events.  In this Letter, we show that the dark matter and metal-poor stars in the Solar neighborhood share similar kinematics due to their common origin.   Using the high-resolution \Eris simulation, which traces the evolution of both the dark matter and baryons in a realistic Milky~Way analog galaxy, we demonstrate that metal-poor stars are indeed effective tracers for the local, virialized dark matter velocity distribution.   The local dark matter velocities can therefore be inferred from observations of the stellar halo made by the Sloan Digital Sky Survey within 4~kpc of the Sun.  This empirical distribution differs from the Standard Halo Model in important ways and suggests that the bounds on the spin-independent scattering cross section may be weakened for dark matter masses below $\sim$10 GeV.  Data from \emph{Gaia} will allow us to further refine the expected distribution for the smooth dark matter component, and to test for the presence of local substructure.

\end{abstract}

\maketitle

\textbf{Introduction.}  The velocity distribution of dark matter (DM) in the Milky Way provides a fossil record of the galaxy's evolutionary history.  In the $\Lambda$CDM paradigm, the Milky Way's DM halo forms from the hierarchical merger of smaller subhalos~\cite{1978ApJ...225..357S}.  As a subhalo falls into, and then orbits, its host galaxy, it is tidally disrupted and continues to shed mass until it completely dissolves.  With time, this tidal debris virializes and becomes smoothly distributed in phase space.  Debris from more recent mergers that has not equilibrated can exhibit spatial or kinematic substructure~\cite{2005ApJ...635..931B,2010MNRAS.406..744C,2011ApJ...733L...7H,2007AJ....134.1579K,2009ApJ...694..130M,2009MNRAS.399.1223S,2010A&ARv..18..567K,Lisanti:2011as,Kuhlen:2012fz,Lisanti:2010qx}.  

Knowledge of the DM velocity distribution is required to interpret results from direct detection experiments~\cite{PhysRevD.31.3059,Drukier:1986tm}, which search for DM particles that scatter off terrestrial targets.  The scattering rate in these experiments depends on both the local number density and velocity of the DM~\cite{1996PhR...267..195J,Freese:2012xd}.  In the Standard Halo Model (SHM), the velocity distribution is modeled as a Maxwell-Boltzmann, which assumes that the DM distribution is isotropic and in equilibrium~\cite{Drukier:1986tm}.  Deviations from these assumptions can be important for certain classes of DM models (see~\cite{Freese:2012xd} for a review).

$N$-body simulations, which trace the build-up of Milky Way--like halos in a cosmological context, do find differences with the SHM.  In DM-only simulations, this is most commonly manifested as an excess of high-velocity particles as compared to a Maxwellian distribution with the same peak velocity~\cite{Vogelsberger:2008qb,MarchRussell:2008dy,Kuhlen:2009vh}.  However, full hydrodynamic simulations, which include gas and stars, find that the presence of baryons makes the DM halos more spherical and the velocities more isotropic, consistent with the SHM~\cite{Ling:2009eh,Kuhlen:2013tra,Bozorgnia:2016ogo, Kelso:2016qqj, Sloane:2016kyi}.

In this Letter, we demonstrate that the DM velocity distribution can be empirically determined using populations of metal-poor stars in the Solar neighborhood.  This proposal relies on the fact that these old stars share a merger history with DM in the $\Lambda$CDM framework, and should therefore exhibit similar kinematics.  The hierarchical formation of DM halos implies that the Milky Way's stellar halo also formed from the accretion, and eventual disruption, of dwarf galaxies~\cite{Searle:1978gc,Johnston:1996sb, Helmi:1999uj, Helmi:1999ks, Bullock:2000qf, Bullock:2005pi}.  For example, the chemical abundance patterns of the stellar halo can be explained by the accretion---nearly 10 Gyr ago---of a few $\sim$$5\times10^{10}$~M$_{\odot}$ DM halos hosting dwarf-irregular galaxies~\cite{Robertson:2005gv, Font:2005qs, Font:2005rm}.  The stars from these accreted galaxies would have characteristic chemical abundances. 

A star's abundance of iron, Fe, and $\alpha$-elements (O, Ca, Mg, Si, Ti) depends on its host galaxy's evolution.  Core-collapse supernova (SN), like Type~II, result in greater $\alpha$-enrichment relative to Fe over the order of a few Myr.  Thermonuclear SN, such as Type~Ia, however, act on longer time scales and produce large amounts of Fe relative to $\alpha$ elements.  For a galaxy that experiences only a brief star-formation period, the enrichment of its interstellar medium is dominated by explosions of core-collapse SN, suppressing Fe abundances.  Observations indicate that the Milky Way's inner stellar halo, which extends out to $\sim$20~kpc, is metal-poor, with an iron abundance of $\text{[Fe/H]}\sim-1.5$ and $\alpha$-enhancement of $\text{[}\alpha\text{/Fe]}\sim0.3$~\cite{1991AJ....101.1835R,1991AJ....101.1865R,1995AJ....109.2757M,2006ApJ...636..804A,2004AJ....128.1177V, Ivezic:2008wk}.\footnote{The stellar abundance of element $X$ relative to $Y$ is defined as:
\begin{equation}
[X/Y] = \log_{10}\left(N_X/N_Y\right) - \log_{10}\left( N_X/ N_Y \right)_\odot \, , \nonumber
\end{equation}
where $N_i$ is the number density of the $i^\text{th}$ element.}  

To demonstrate the correlation between the stellar and DM velocity distributions, we use the \textsc{Eris} simulation, one of the highest resolution hydrodynamic simulations of a Milky Way--like galaxy~\cite{Guedes:2011ux}.  We show that the velocity distribution of metal-poor halo stars in \textsc{Eris} successfully traces that of the virialized DM component in the Solar neighborhood.  The correspondence between the DM and stellar kinematics is best when $\FeH < -3$.  In general, the average metallicity of the stars in a satellite correlates with its mass~\cite{Kirby:2013wna}, with the most metal-poor stars dominating in the ultrafaint dwarf galaxies with stellar masses below $\sim$$10^5$~M$_\odot$.  While   the metallicity distributions of classical dwarfs with stellar masses between $\sim$$10^5$--$10^8$~M$_\odot$ typically have average values $\FeH \gtrsim -2$, their tails extend down to much lower values.  As a result, these satellites can contribute a significant fraction ($\sim$40--80\%) of the stars with $\FeH < -2$, as was shown in~\cite{Deason:2016wld}.  Because the very metal-poor end of the stellar distribution samples tidal debris from a broad swath of satellite masses, it does a better job at tracing the kinematics of the dark matter halo, which is a product of the full merger history.  

Given the correlation observed in \textsc{Eris}, we then use results from the Sloan Digital Sky Survey (SDSS) to infer the local velocity distribution for the smooth DM component in our Galaxy.  The result differs from the SHM in important ways.  If these results continue to hold as upcoming surveys probe increasingly lower metallicities, it would suggest that current limits on spin-independent DM may be too strong for masses below $\sim$10~GeV. 

\textbf{The Eris Simulation.}  \textsc{Eris} is a cosmological zoom-in simulation that employs smoothed particle hydrodynamics to model the DM, gas, and stellar distributions in a Milky Way--like galaxy from $z=90$ to today~\cite{Guedes:2011ux, Guedes:2012gy}.  It employs the TreeSPH code \textsc{Gasoline}~\cite{Wadsley:2003vm} to simulate the evolution of the galaxy in a WMAP cosmology~\cite{Spergel:2006hy}.  The mass resolution is $9.8\times10^4$ and $2\times10^4$~M$_\odot$ for each DM and gas `particle,' respectively.  An overview of the simulation is provided in~Refs. \cite{Guedes:2011ux, Guedes:2012gy,Pillepich:2014jfa,2016arXiv161202832S,2015ApJ...807..115S}, and we summarize the relevant aspects for our study here.  

The \textsc{Eris} DM halo has a virial mass of $M_\text{vir} = 7.9\times10^{11}$~M$_\odot$ and radius $R_\text{vir} = 239$~kpc, and experienced no major mergers after $z=3$.  Within $R_\text{vir}$, there are $7\times10^6$, $3\times10^6$, and $8.6\times10^6$ DM, gas, and star particles, respectively.  At $z=0$, the DM halo hosts a late-type spiral galaxy.  The disk has a scale length of 2.5~kpc and exponential scale height of 490~pc at 8~kpc from the galactic center.  The properties of the \Eris disk and halo are comparable to their Milky Way values~\cite{Guedes:2011ux, Pillepich:2014jfa}

A star `particle' of mass $6\times10^3$~M$_\odot$ is produced if the local gas density exceeds $5$ atoms/cm$^3$.  The star formation rate depends on the gas density, $\rho_\text{gas}$, as $d\rho_*/dt = 0.1\, \rho_\text{gas}/ t_\text{dyn} \propto \rho_\text{gas}^{1.5}$, where $\rho_*$ is the stellar density and $t_\text{dyn}$ is the dynamical time.
Metals are redistributed by stellar winds and Type~Ia and Type~II SNe~\cite{2016arXiv161202832S,2015ApJ...807..115S}.  The abundances of Fe and O are tracked as the simulation evolves, while the abundances of all other elements are extrapolated assuming their measured solar values~\cite{Asplund:2009fu}.  The Supplementary Material provides a detailed explanation for the chemical abundance modeling in \textsc{Eris}.  

Stars may either be bound to the main host halo or to its satellites when they form.  We are primarily interested in the latter, as these stars share a common origin with the DM.  The vast majority of halo stars in \Eris originated in satellites and are older than those born in the host~\cite{Pillepich:2014jfa}.  They are more metal-poor than disk stars, on average, and we take advantage of this difference to distinguish the two components in the \textsc{Eris} galaxy.  

\begin{figure}[t]
\begin{center}
\includegraphics[width=0.45\textwidth]{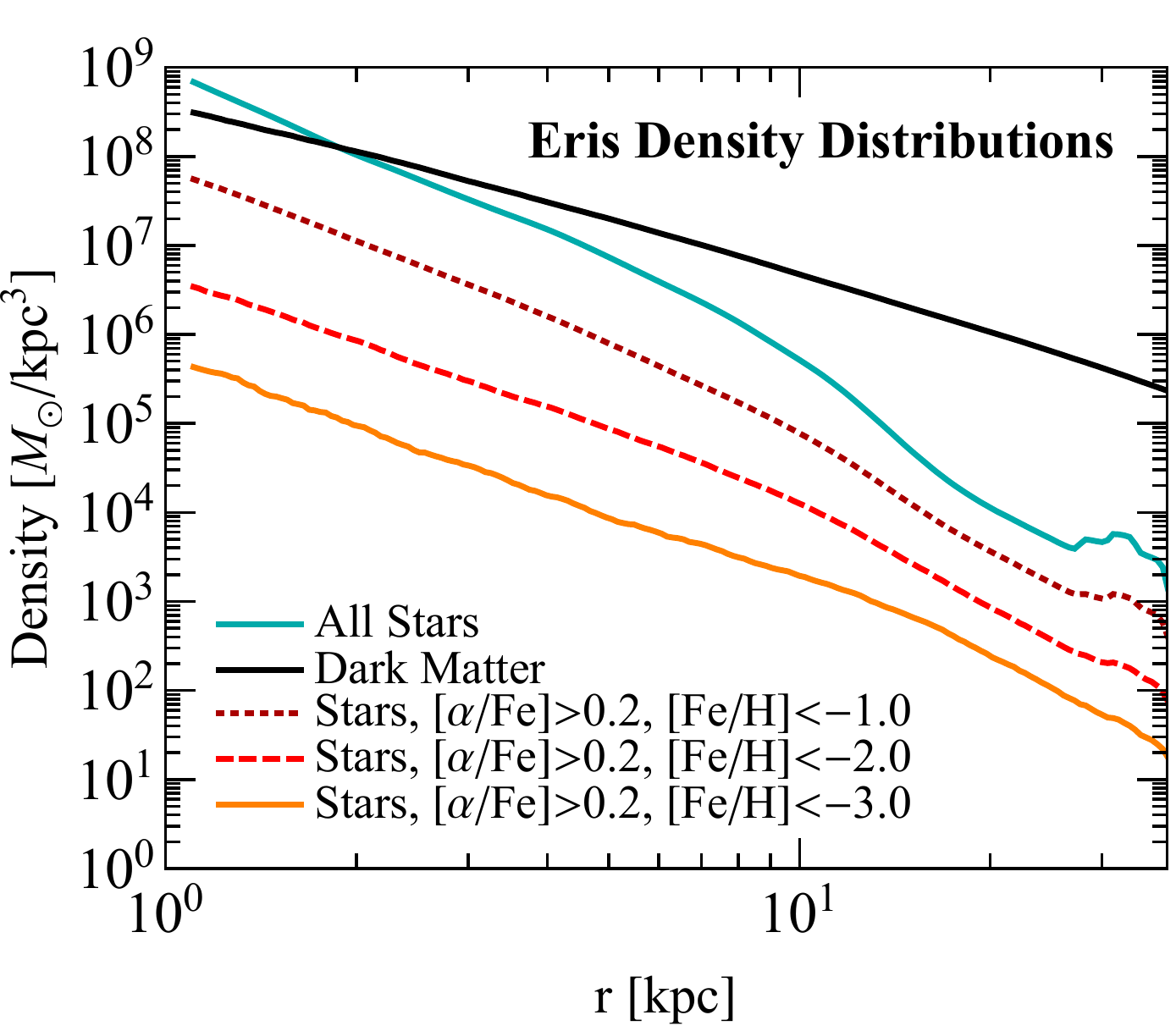}
\end{center}
\vspace{-0.2in}
\caption{The density distribution as a function of Galactocentric radius for the dark matter (black) and all stars (cyan) in \textsc{Eris}.  The distributions for subsamples of stars with $\alphaFe> 0.2$ and $\FeH <-1,-2,-3$ are also shown (dotted brown, dashed red, and solid orange, respectively).  The density of the most metal-poor stellar population exhibits the same dependence on radius as the dark matter near the Sun's position, $r_\odot\sim8$~kpc.}
\label{fig:rho}
\vspace{-0.2in}
\end{figure}

\textbf{Stellar Tracers for Dark Matter.}  Figure~\ref{fig:rho} shows the density distribution of the DM and stars in \textsc{Eris} as a function of Galactocentric radius.  The distribution for all stars is steeper than that for DM.  However, this includes contributions from thin and thick disk, as well as halo stars.  To select the stars that are most likely to be members of the halo, we place cuts on both the Fe and $\alpha$-element abundances.  Figure~\ref{fig:rho} illustrates what happens when progressively stronger cuts are placed on [Fe/H], while keeping $\alphaFe > 0.2$.  As the cut on iron abundance varies from $\FeH < -1$ to $\FeH < -3$, the density fall-off becomes noticeably more shallow.  
\begin{figure*}[t]
\begin{center}
\includegraphics[width=0.34\textwidth]{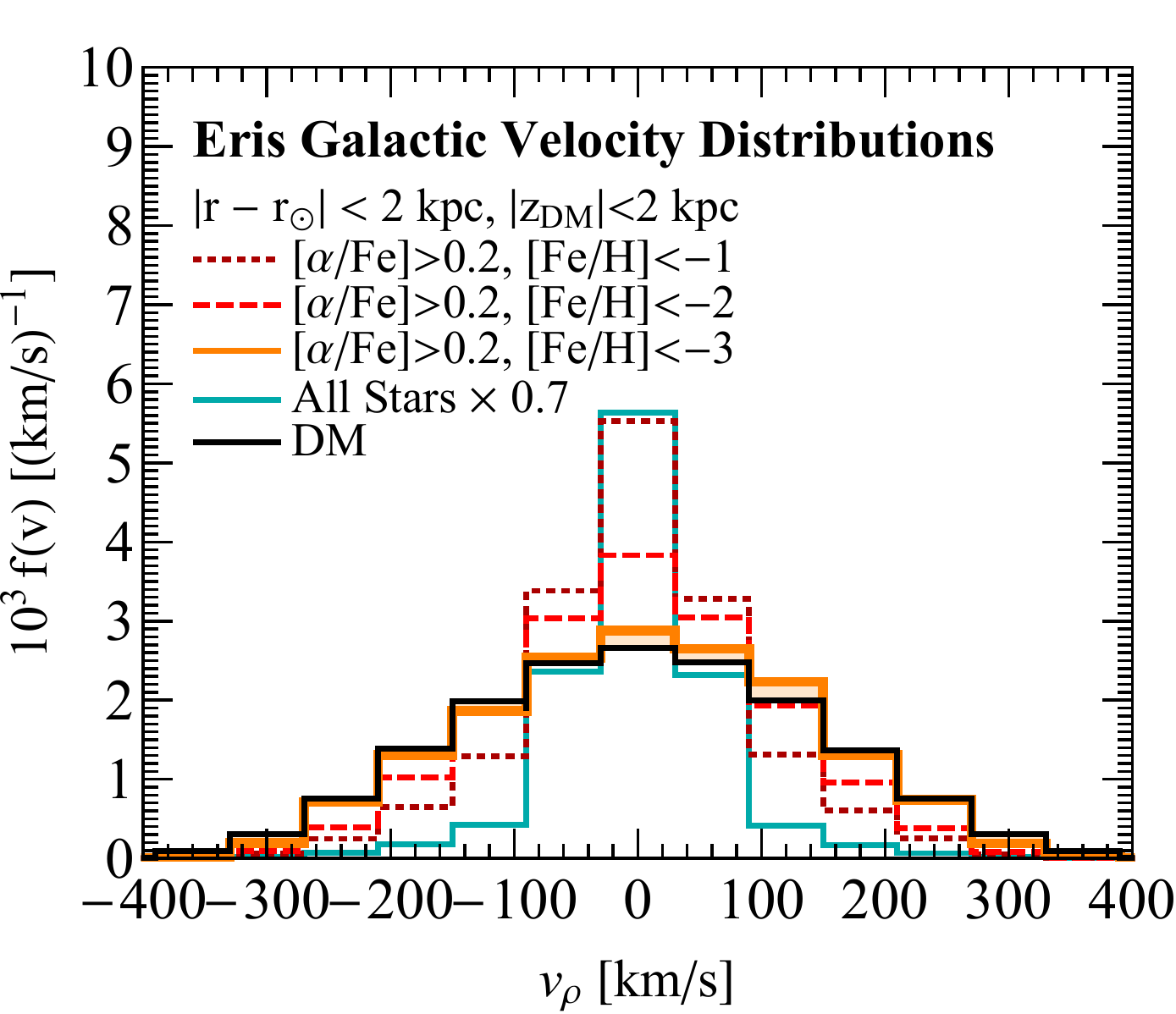}
\includegraphics[trim={0.8cm 0 0 0},clip,width=0.32\textwidth]{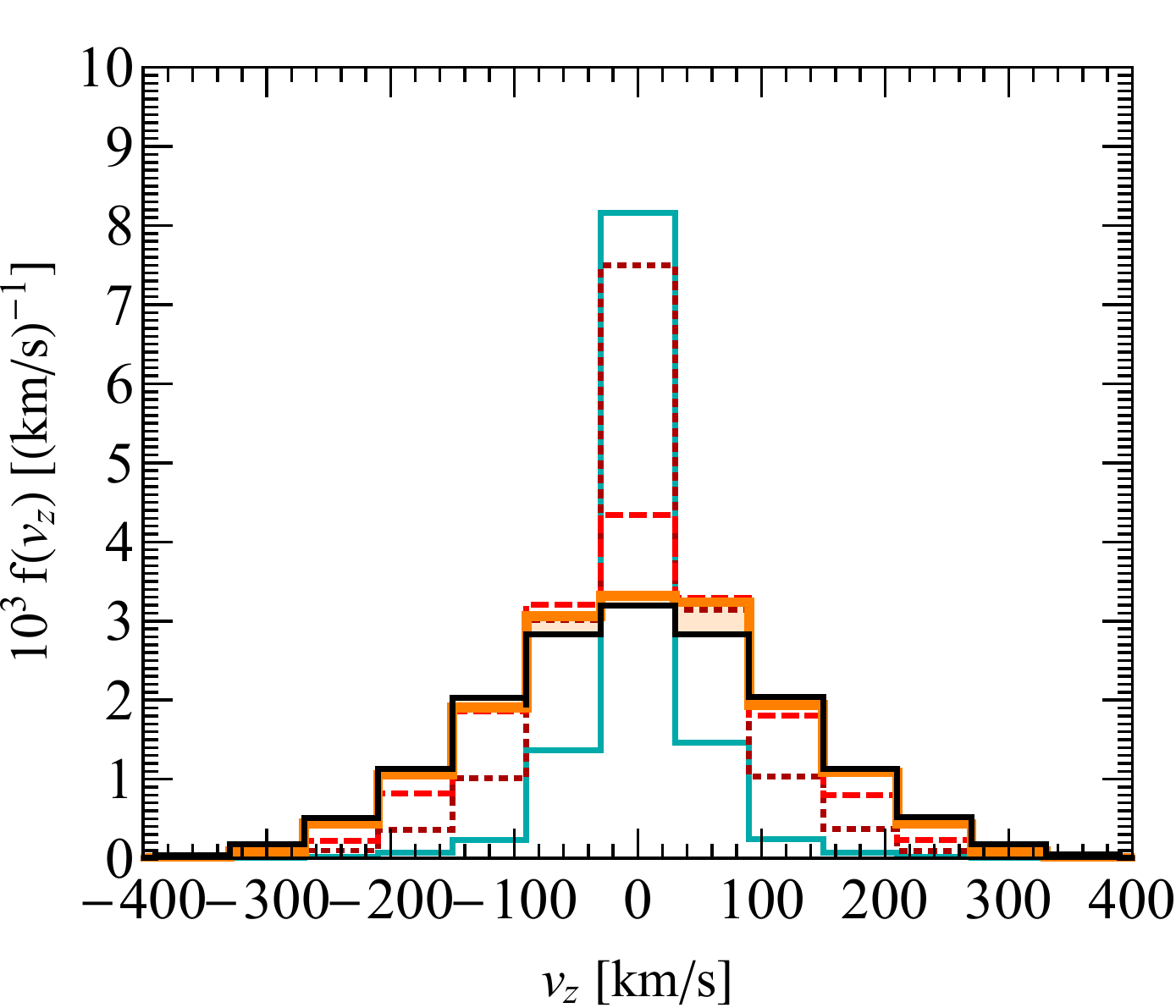}
\includegraphics[trim={0.8cm 0 0 0},clip,width=0.32\textwidth]{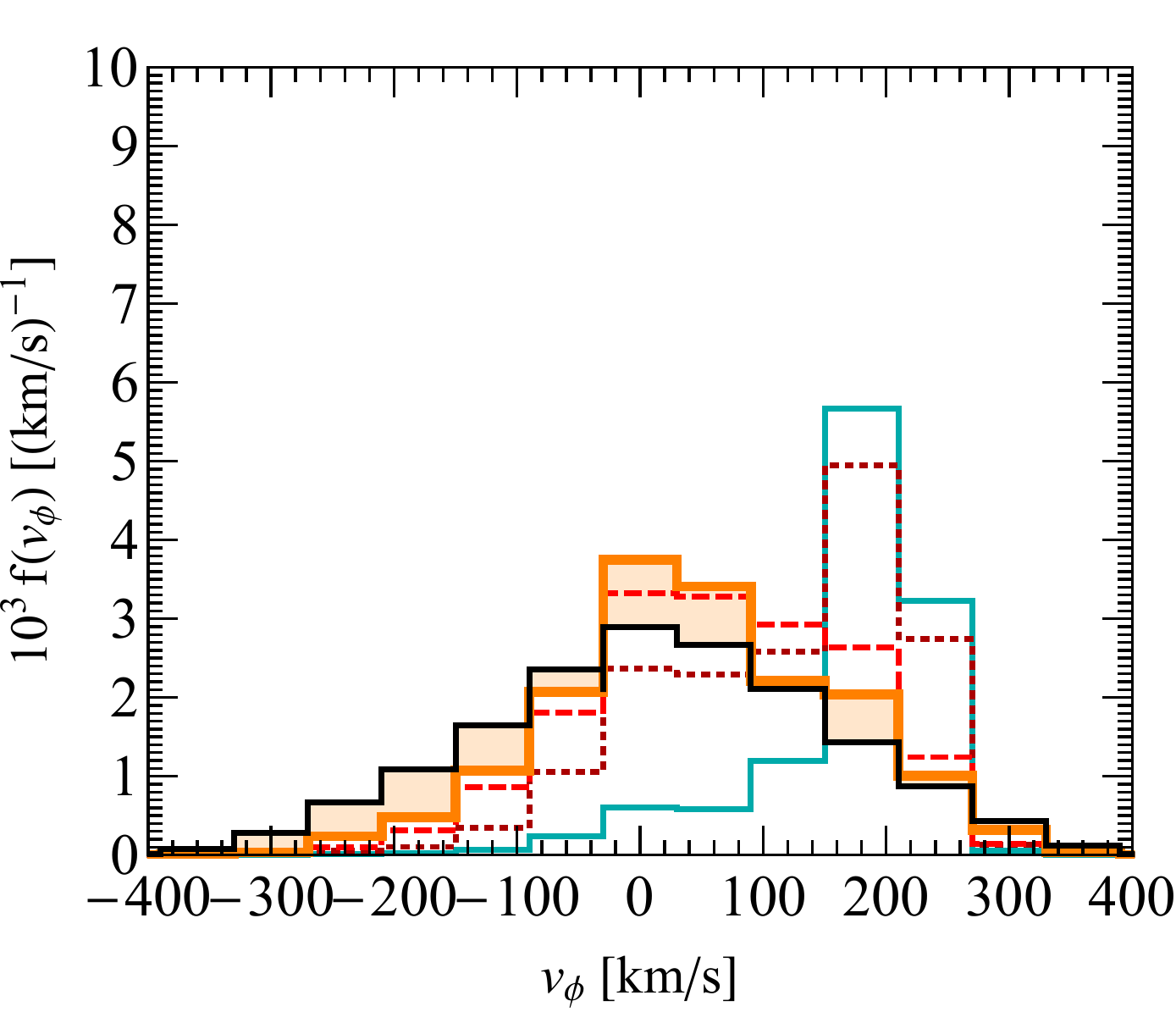}
\end{center}
\vspace{-0.2in}
\caption{Distributions of the three separate velocity components of the DM (solid black) and stars in \textsc{Eris}. The velocities are in the galactocentric frame, where the $z$-axis is oriented along the stellar angular momentum vector. The stellar distributions are shown separately for different metallicities, with $\alphaFe >0.2$ and iron abundance varying from $\FeH < -1$ (dotted brown) to $\FeH < -3$ (solid orange).  The distribution for all stars---dominated primarily by the disk---is also shown (solid cyan).  All distributions are shown for $|r - r_\odot| \leq 2$~kpc; the DM is additionally required to lie within 2~kpc of the plane.  To guide the eye, the orange shading highlights the differences between the DM and $\FeH < -3$ distributions.  The discrepancy in the $v_\phi$ distributions is due to the preferential disruption of subhalos on prograde orbits in \textsc{Eris}; observations of the Milky Way halo do not see such pronounced prograde rotation~\cite{Carollo:2007xh,Bond:2009mh}. }
\label{fig:bestfit}
\vspace{-0.2in}
\end{figure*}

Because the focus of this work is the DM distribution in the Solar neighborhood, we consider galactocentric radii in the range $|r - r_\odot| \leq 2$~kpc, where $r_\odot = 8$~kpc is the Sun's position.  In this range, the DM distribution falls off as $\rho(r) \propto r^{-2.07 \pm 0.01}$, which is essentially consistent with the best-fit power-law for the most metal-poor subsample, which falls off as $\rho(r) \propto r^{-2.24 \pm 0.12}$. This illustrates that the stars with lower iron abundance are adequate tracers for the underlying DM density distribution (see also~\Ref{Tissera:2014uwa}).  The correspondence between the density distributions breaks down above $r \gtrsim20$~kpc, indicating a transition from the inner to the outer halo that is consistent with observations~\cite{Carollo:2007xh}.

Figure~\ref{fig:bestfit} compares the velocity distribution of candidate halo stars in \Eris with that of the DM.\footnote{Throughout, we define the $z$-axis to be oriented along the angular momentum vector of the stars.}  For comparison, we also show the stellar distribution with no metallicity cuts; it is dominated by disk stars with a characteristic peak at $v_\phi \simeq 220$~km/s and narrow dispersions in the radial and vertical directions.  All distributions are shown for $|r - r_\odot| \leq 2$~kpc.  Because direct detection experiments are only sensitive to DM within the Solar neighborhood, we restrict its vertical displacement from the disk to be $|z_\text{DM}| \leq 2$~kpc.  The stellar distributions are shown with no cut on the vertical displacement---that is, with only the $|r - r_\odot| \leq 2$~kpc requirement applied.  The stellar distributions become statistics-limited if $|z| \leq z_0$ for $z_0 = 2$~kpc is also required.  We have verified that the results do not change if we restrict the metal-poor population to vertical displacements where $z_0 > 2$~kpc.  

The $v_\rho$ and $v_z$ distributions show an excellent correspondence between the halo stars and the DM.  Indeed, as increasingly more metal-poor stars are selected, their velocity distribution approaches that of the DM exactly.  We apply the two-sided Kolmogorov-Smirnov test to establish whether the DM and halo stars share the same $v_\rho$ and $v_z$ probability distributions.  The null hypothesis that the DM and stars share the same parent distribution is rejected at 95\% confidence if the $p$-value is less than $0.05$.  The $p$-values for the $(v_\rho, v_z)$ distributions are $(0.9, 0.1)$ for $\FeH < -3$, suggesting that its velocity distribution is indistinguishable from that of the DM in the radial and vertical directions. 

Interpreting the distribution of azimuthal velocities requires more care.  As illustrated in Fig.~\ref{fig:bestfit}, the azimuthal velocities are skewed to positive values for both the DM and halo stars.  The prograde rotation in the DM distribution is attributable to the `dark disk,' which comprises $\sim$9\% of all the DM in the Solar neighborhood in \textsc{Eris}~\cite{Pillepich:2014jfa}.  Dark disks form from the disruption of subhalos as they pass through the galactic disk.  Subhalos on prograde orbits are preferentially disrupted due to dynamical friction, leading to a co-rotating DM disk~\cite{Read:2008fh}.  The effect on the stars is similar, and---indeed---more pronounced due to dissipative interactions between halo stars and the disk~\cite{Pillepich:2014jfa}.  The end result is that the halo stars systematically under-predict the DM distribution at negative azimuthal velocities.  

Current observations suggest that our own Milky Way has an inner halo with either modest or vanishing prograde rotation~\cite{Carollo:2007xh,Bond:2009mh}, and constrain the possible contributions from a dark disk~\cite{Read:2014qva}.  This suggests that the mergers that resulted in \textsc{Eris}' prograde halo might not have occurred in our own Galaxy, making the comparison of the DM and halo azimuthal motions more straightforward in realization.  In the absence of such mergers, we assume that the DM and metal-poor stars have $v_\phi$ distributions that match just as well as those in the $v_\rho$ and $v_z$ cases.

We have verified that the results presented in Fig.~\ref{fig:bestfit} are robust even as the spatial and \alphaFe cuts are varied.  We consider $\alphaFe \in [0.2, 0.4]$, remove the \alphaFe cut altogether, and study the region where $|r-r_\odot| \leq 1$~kpc.  In all these cases, the conclusions remain the same.
  
\textbf{Empirical Velocity Distribution.}  We now look to the kinematic properties of the Milky Way's stellar halo to infer the local DM velocities by extrapolating the correspondence argued above to our Galaxy.  We use as a reference the sample studied in~\cite{2010ApJ...712..692C}, where spatial, chemical, and kinematic properties of halo stars have been characterized using SDSS data.
The sample includes stars within distances of 4~kpc of the Sun and Galactocentric radii $7 < r <10$~kpc.  
The cylindrical velocity components for the stars in this volume are provided in separate metallicity bins and we focus on the lowest two: $\FeH \in [-2.2, -2]$ and $< -2.2$.  

\begin{figure}[t]
\begin{center}
\includegraphics[width=0.45\textwidth]{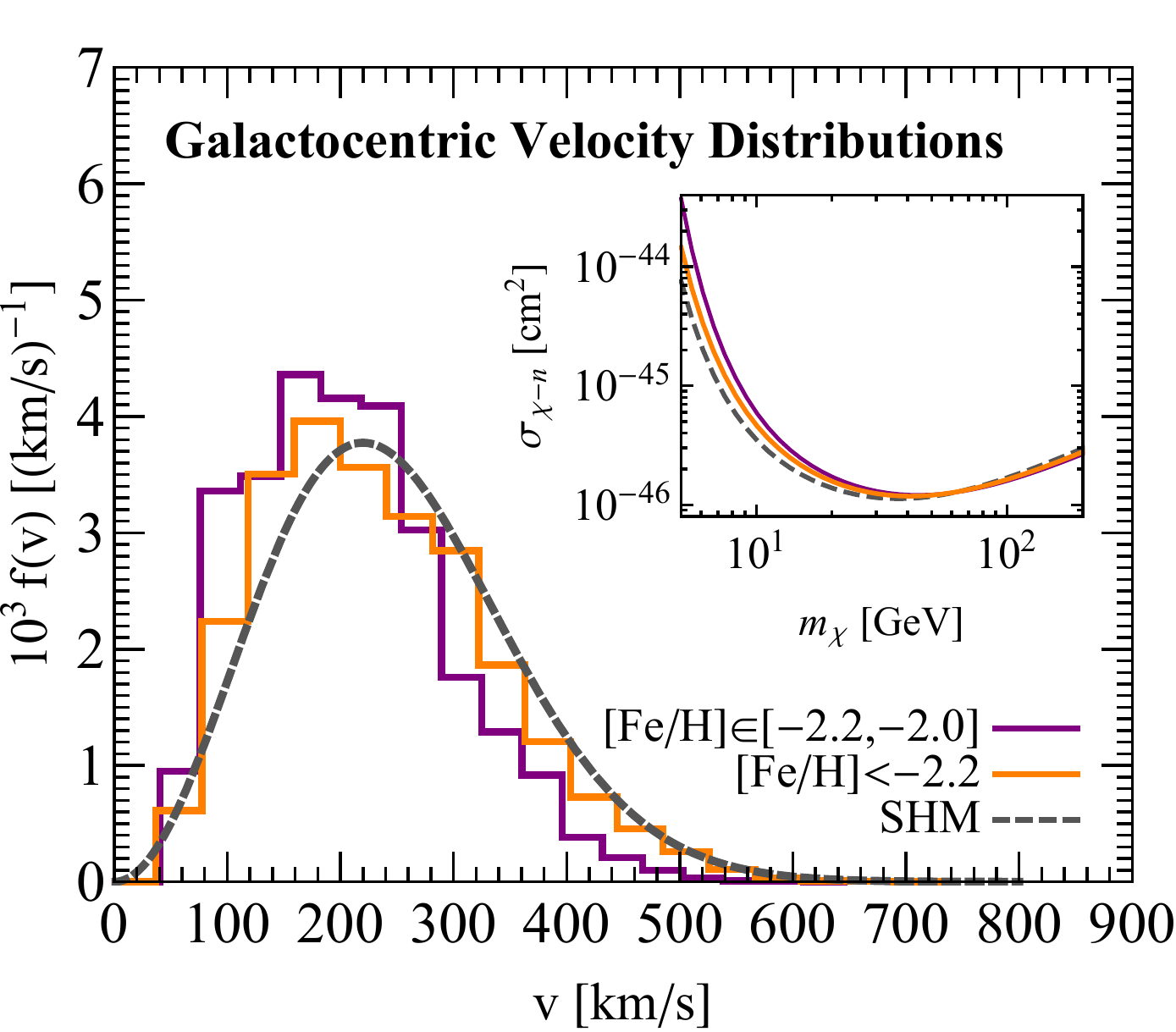}
\end{center}
\vspace{-0.2in}
\caption{Galactocentric speed distributions for SDSS stars within 4~kpc of the Sun and Galactocentric distances of $7 < r < 10$~kpc, based on results from~\cite{2010ApJ...712..692C}.  The distributions are shown for $\FeH \in [-2.2, -2]$~(solid purple) and $\FeH < -2.2$~(solid orange).  For comparison, we show the Standard Halo Model (dashed gray) with $v_c = 220$~km/s.  Not captured by this figure is the fact that the stellar distributions are not isotropic, as is typically assumed for the Standard Halo Model.  The inset shows the expected background-free 95\% C.L. limit on the DM spin-independent scattering cross section, assuming the exposure and energy threshold of the LUX experiment~\cite{Akerib:2016vxi} for the SDSS and SHM velocity distributions.}
\label{fig:sdss}
\vspace{-0.2in}
\end{figure}

Figure~\ref{fig:sdss} shows the Galactocentric speed distributions for the metal-poor stars, obtained by generating a  mock catalog from the distributions of the separate velocity components provided in~\cite{2010ApJ...712..692C}.  We assume that the velocity components are uncorrelated, which is only approximately valid; small $\mathcal{O}(10^\circ)$ correlations have been observed between the radial and z-component cylindrical velocities~\cite{2010ApJ...712..692C,Bond:2009mh}.  
The distribution for $\FeH < -2.2$ has a larger dispersion than that for $\FeH \in [-2.2, -2]$.  The peak speed for both distributions is essentially the same, falling within the range $v\sim$160--180~km/s.

For comparison, the SHM is also shown in Fig.~\ref{fig:sdss}.  The SHM has isotropic dispersions ($\sigma = \sigma_{r,\phi,z}$) and is described by a Maxwell-Boltzmann distribution $f(v) \propto e^{-v^2/2\sigma^2}$.  This corresponds to a collisionless isothermal distribution with density $\rho\left(r\right) \propto r^{-2}$, and yields a flat rotation curve with circular velocity $v_c^2 = 2 \sigma^2$, where $v_c \sim 220$~km/s.\footnote{Recent determinations of the disk rotation speed typically place its value higher by $\sim$5--15\%, though some models predict best-fit values as low as $\sim$200~km/s---see \cite{Freese:2012xd} and references therein.}  We see that the stellar speed distributions approach that of the  SHM as more metal-poor stars are selected, although they systematically underestimate the number of high-speed particles above $v \gtrsim 200$~km/s.  We stress that there is an important difference between the SHM and SDSS distributions that is not evident from Fig.~\ref{fig:sdss}: namely, the metal-poor stellar distributions are not isotropic, a basic assumption underlying the SHM.

If the SDSS halo stars are adequate tracers for the local DM, then Fig.~\ref{fig:sdss} suggests that the DM speeds may be slower, on average, than what is expected in the SHM. This can lead to noticeable differences in the predicted signal rate for direct detection experiments.  If a DM particle of mass $m_\chi$ scatters off a nucleus with momentum transfer $q$ and effective cross section $\sigma(q^2)$, the scattering rate is
\begin{equation}
\frac{dR}{dE_\text{nr}} = \frac{\rho_\chi}{2 m_\chi \mu} \sigma(q^2) F(q) \, \int_{v_\text{min}}^\infty \frac{f \left(\textbf{v} + \textbf{v}_\text{obs}(t)\right)}{v} \, d^3 v \, ,
\end{equation}
where $E_\text{nr}$ is the recoil energy of the nucleus, $\rho_\chi$ is the local DM density, $\mu$ is the DM-nucleus reduced mass, $F(q)$ is the exponential nuclear form factor \cite{1996PhR...267..195J}, $v_\text{min}$ is the minimum velocity needed to scatter, and $\textbf{v}_\text{obs}(t)$ is the velocity of the lab frame relative to the Galactic frame.\footnote{When transforming the stellar speed distributions in Fig.~\ref{fig:sdss} to the heliocentric frame, we assume that they are spatially isotropic.}  Taking the exposure of the LUX experiment, with $3.35\times 10^{4}$ kg days and a minimum energy threshold of $1.1$ keV \cite{Akerib:2016vxi}, we derive the 95\% one-sided Poisson C.L bound (3.0 events) on the scattering cross section as a function of the DM mass.  The result is shown in the inset of \Fig{fig:sdss} for the SHM and SDSS distributions.  The bounds on the lightest DM are weakened when the empirical distribution is used rather than the SHM.

\textbf{Conclusions.}  In this Letter, we propose that DM velocities can be determined empirically using metal-poor stars in the Solar neighborhood.  Low metallicity stars are typically born in galaxies outside our own.  Like DM, they are dragged into the Milky Way through mergers, and predominantly populate the halo surrounding the disk.  We demonstrate the close correlation between the distributions of DM and metal-poor stars using the \textsc{Eris} simulation, and conclude that the kinematics of the stellar population tracks that of the virialized DM.  To verify the generality of these findings and understand their dependence on the merger history, this study should be repeated with other hydrodynamic simulations of Milky Way--like halos and generalizations of $\Lambda$CDM, such as self-interacting DM.  In future work, we  plan to determine the origin of the metal-poor stars in \textsc{Eris} and characterize the properties of the merger history that dictate the correspondence between the stars and the dark matter.  This will strengthen our understanding of how to generalize the simulation results to our own Galaxy.  

This Letter focused on the virialized DM component, but we now comment briefly on recovering information regarding DM substructure.  The velocity distribution of the most metal-poor stars should predominantly sample the oldest mergers~\cite{Deason:2016wld} and therefore correlate with the virialized DM.  In contrast, substructure should arise from younger mergers whose tidal debris has not fully virialized with the host.  Whether these recently disrupted satellites contribute stellar debris at the very metal-poor end of the spectrum depends on the details of their respective metallicity distributions.  If they do contribute, then one way to potentially separate their contributions from the virialized component is to identify outliers in the velocity distribution of the most metal-poor stars.  Such outliers could point to phase-space substructure as the debris from recent mergers is typically concentrated at the highest velocities~\cite{Lisanti:2011as,Lisanti:2014dva,Kuhlen:2009vh}.  Current evidence from both observation~\cite{Kuhlen:2009vh} and simulation~\cite{2011MNRAS.413.1419V} suggests that the Solar neighborhood is not dominated by a single stream, however this does not preclude the possibility of lower-density streams or debris flows.

As a first step towards characterizing the DM velocity distribution empirically, we used published results of the velocity distribution of the most metal-poor stars within 4~kpc of the Sun, obtained from SDSS~\cite{2010ApJ...712..692C}.  The corresponding speed distribution for the most metal-poor of the stars sampled has a lower peak speed and smaller dispersion than what is typically assumed in the SHM.  In addition, the total velocity distribution is not isotropic, as assumed for the SHM.  If this trend continues to hold to lower metallicities, it may affect predictions for the DM scattering rate in direct detection experiments, weakening the published limits on the spin-independent cross section at masses below $\sim$10~GeV.   
The wealth of data from \emph{Gaia}~\cite{Perryman2001} allows us to expand upon the SDSS results.  In a follow-up paper, we utilize the first \emph{Gaia} data release and perform a full statistical comparison of the local metal-poor stellar halo with the SHM expectation, study the consequences for direct detection, and test for the possibility of additional substructure from recent mergers~\cite{ravepaper}.  

\textbf{Acknowledgements.} We thank George Brova, Anna Frebel, Nick Gnedin, David Hogg,  Alexander Ji, Mario~Juri$\displaystyle{\acute{\text{c}}}$, Annika Peter, Adrian Price-Whelan, and $\displaystyle{\check{\text{Z}}}$eljko~Ivezi$\displaystyle{\acute{\text{c}}}$ for helpful conversations.  M.L. is supported by the DOE under contract DESC0007968, as well as by the Alfred P.~Sloan Foundation.  P.M. thanks the Pr\'{e}fecture of the Ile-de-France Region through the award of a Blaise Pascal International Research Chair, managed by the Fondation de l'Ecole Normale Sup\'{e}rieure.  L.N. is supported by the DOE under contract DESC00012567.

\bibliography{eris}{}
\newpage

\onecolumngrid

\newpage
\appendix
\setcounter{equation}{0}
\setcounter{figure}{0}
\setcounter{table}{0}
\setcounter{section}{0}
\makeatletter
\renewcommand{\theequation}{S\arabic{equation}}
\renewcommand{\thefigure}{S\arabic{figure}}
\renewcommand{\thetable}{S\arabic{table}}

\section{Supplementary Material}

This Supplementary Material reviews how the chemical abundances are determined in the \textsc{Eris} simulation~\cite{Guedes:2011ux, Guedes:2012gy,2016arXiv161202832S,2015ApJ...807..115S}.  The Fe and O yields depend on both the initial mass function (IMF), which sets the number of Type Ia and II supernovae (SNe), as well as the assumed chemical yields from each.  The \textsc{Eris} yields are obtained following a procedure similar to that of~\Ref{1996A&A...315..105R}.  Each $M_s=6\times10^3$~M$_\odot$ stellar particle in \textsc{Eris} represents its own population of stars.  The distribution of masses for the particle's constituents are set by the Kroupa~(1993) IMF~\cite{Kroupa:1993ga}:
\begin{equation}
\Phi_\text{Kroupa}(m) = M_s \times N_K \times \left\{
        \begin{array}{ll}
            2^{0.9}\, m^{-1.3} & \quad 0.08 \leq m < 0.5 \, \text{M}_\odot \\
            m^{-2.2} & \quad 0.5 \leq m < 1\, \text{M}_\odot \\
            m^{-2.7} & \quad m \geq 1\, \text{M}_\odot  \, ,
        \end{array}
    \right.
    \label{eq:Kroupa}
\end{equation}
where $m$ is the stellar mass in units of M$_\odot$ and $N_K= 0.303$ is a normalization constant chosen such that $\int m \, \Phi_\text{Kroupa} \text{d}m = M_s$~M$_\odot$ for $0.08\leq m\leq 100$~M$_\odot$.  

Stars with masses between 8 and 40~M$_\odot$ explode as Type~II SN, depositing iron and oxygen with mass yields
\begin{eqnarray}
M_\text{Fe} &=&  2.802\times10^{-4} \, m^{1.864} \, \text{M}_\odot \nonumber \\
M_\text{O} &=& 4.586\times 10^{-4} \, m^{2.721}\, \text{M}_\odot \, ,
\label{eq:Raiteri}
\end{eqnarray} 
which follow the parametrization of~\Ref{1995ApJS..101..181W}.  Determining the number of Type~Ia SN that are produced is more involved---see~\Ref{1996A&A...315..105R} for a full description.  Each Type~Ia SN produces 0.63~M$_\odot$ and 0.13~M$_\odot$ of Fe and O, respectively~\cite{1986A&A...158...17T}.  

Using the IMF and yields described above, we find that a single star particle in \textsc{Eris} produces 29.2 and 4.68 Type~II and Ia SNe, respectively.  This results in 1.32~M$_\odot$ of Fe and 28.5~M$_\odot$ of O from Type~II SN, as well as 2.95~M$_\odot$ of Fe and 0.608~M$_\odot$ of O from Type~Ia SN.  

\begin{figure}[p]
\begin{center}
\includegraphics[width=0.4\textwidth]{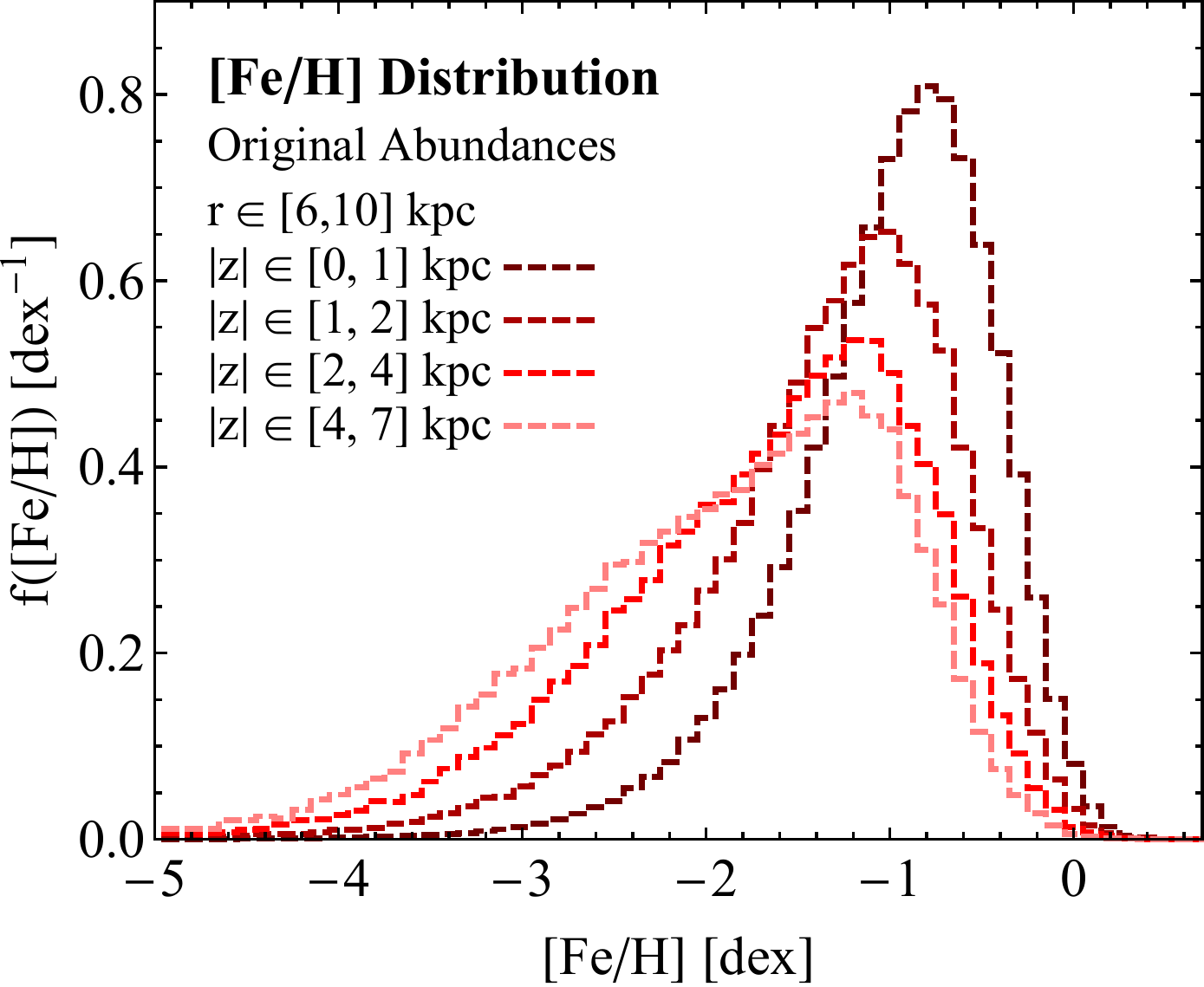}
\qquad
\includegraphics[width=0.4\textwidth]{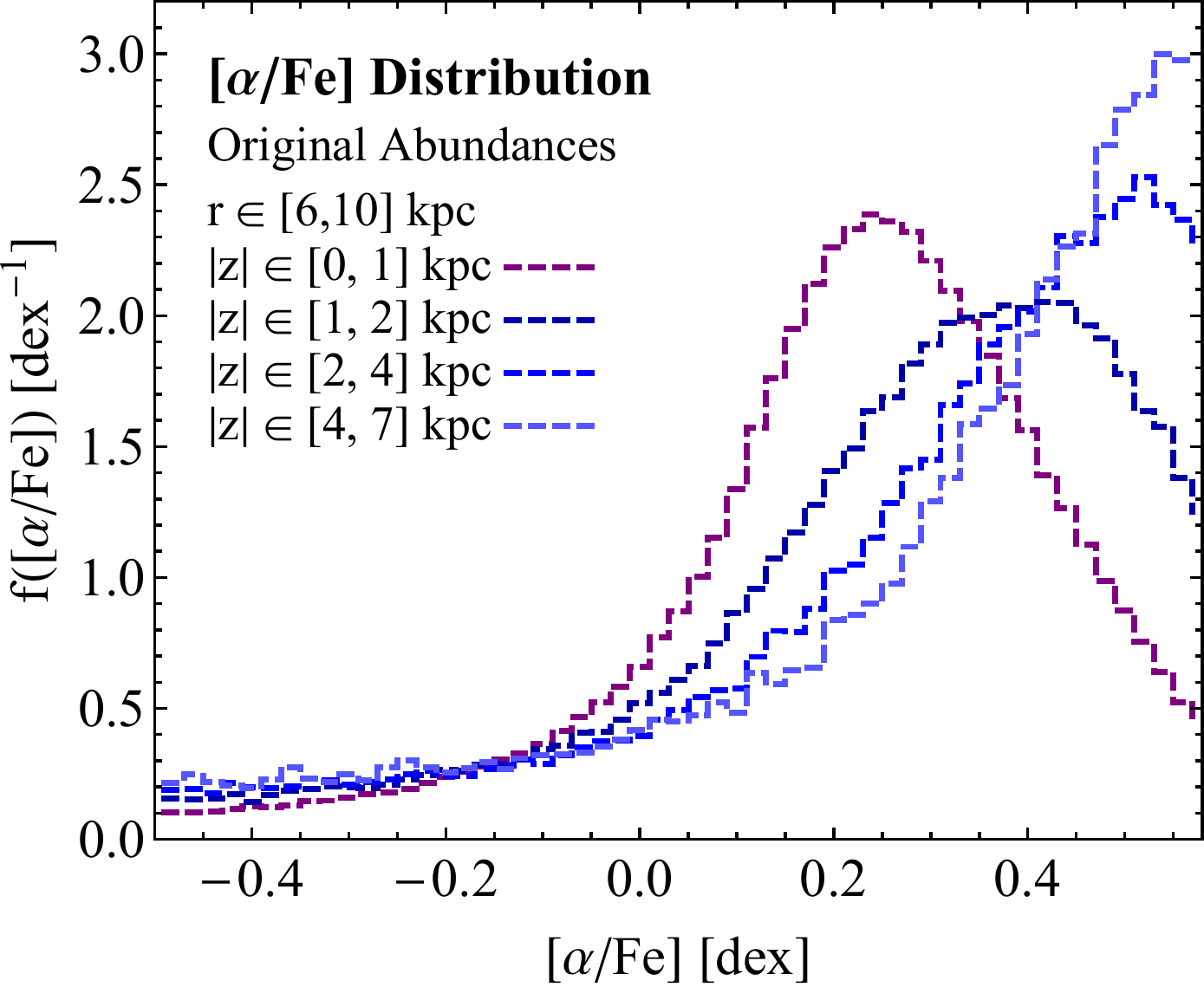}
\includegraphics[width=0.4\textwidth]{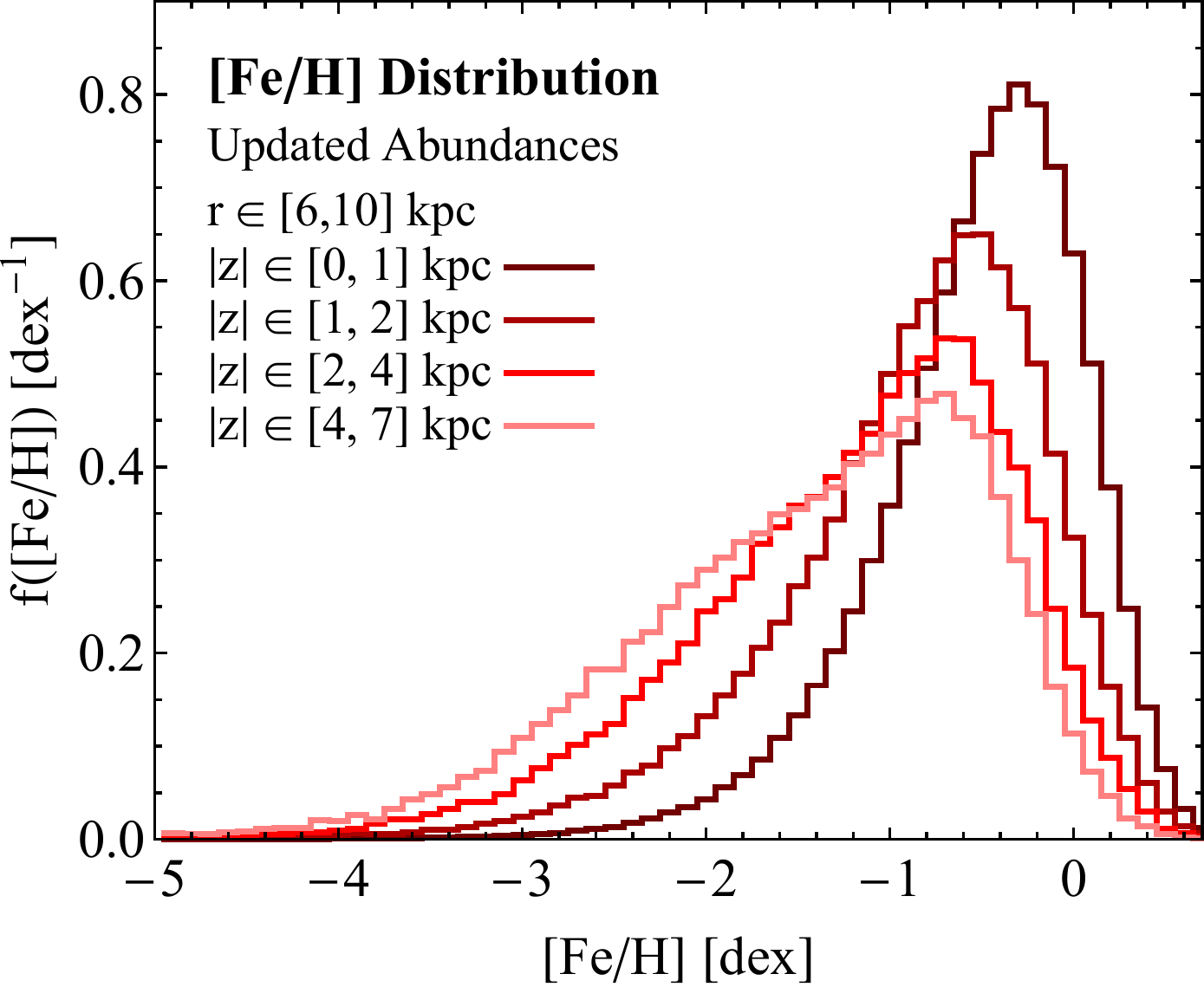}
\qquad
\includegraphics[width=0.4\textwidth]{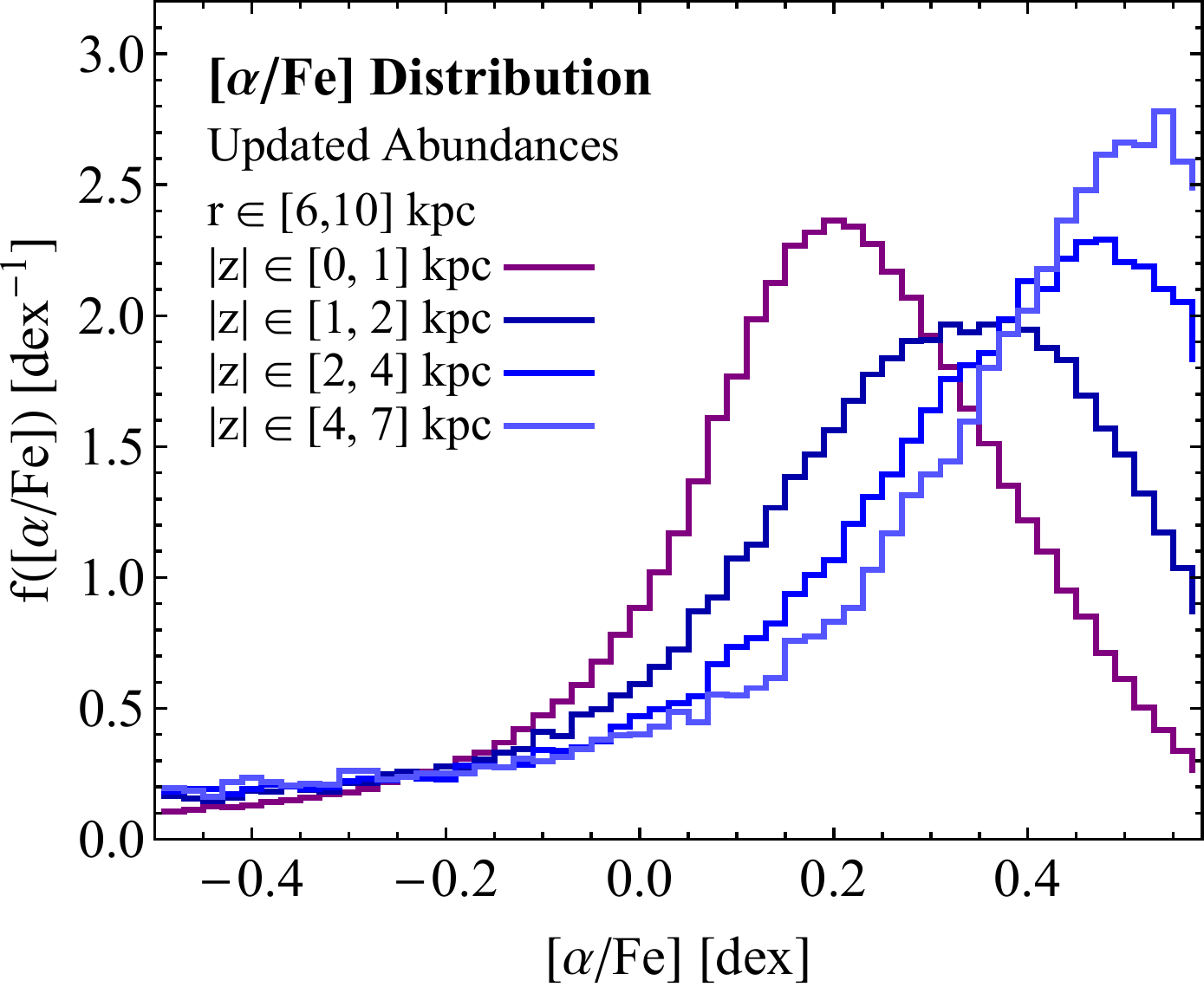}
\includegraphics[width=0.4\textwidth]{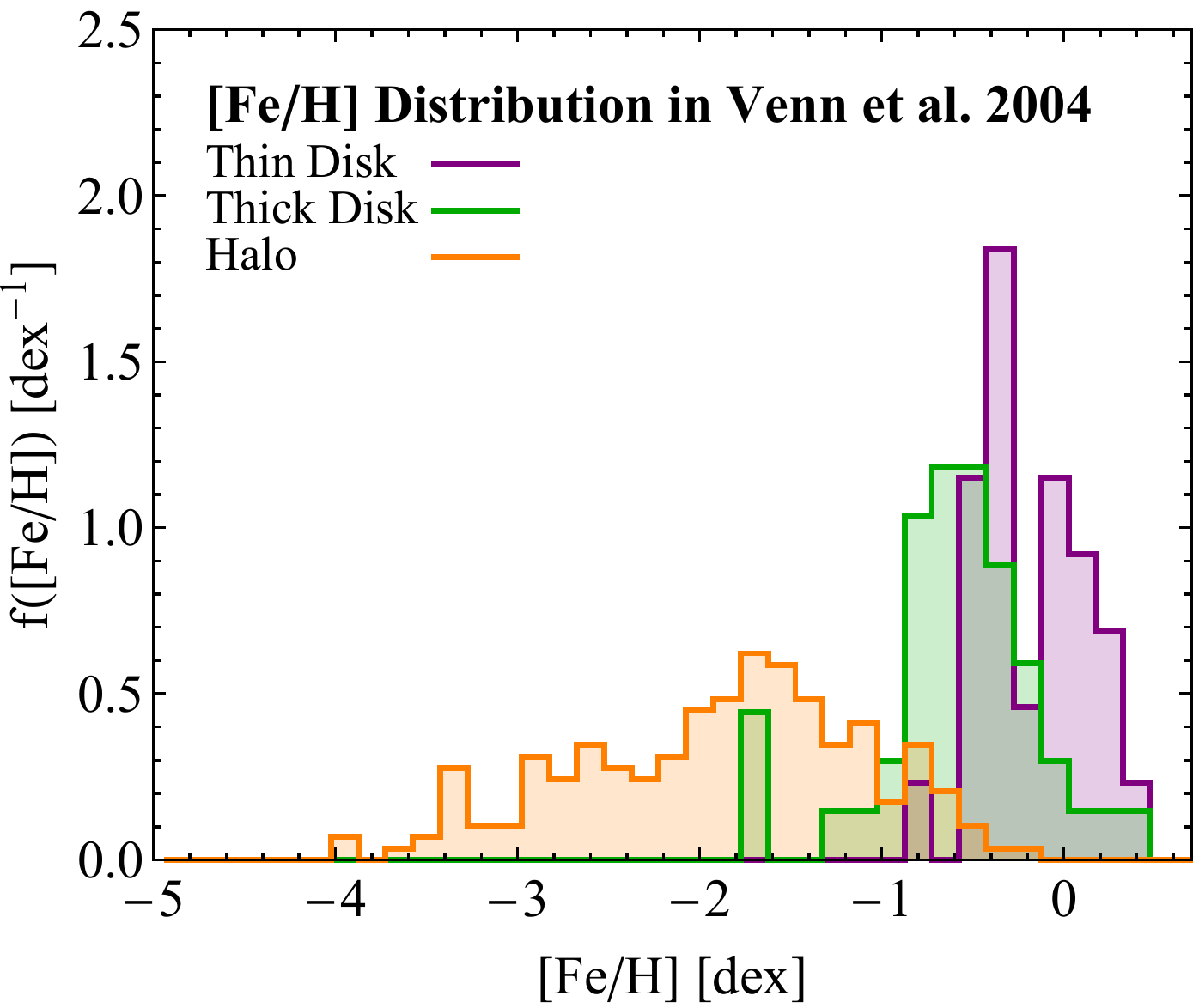}
\qquad
\includegraphics[width=0.4\textwidth]{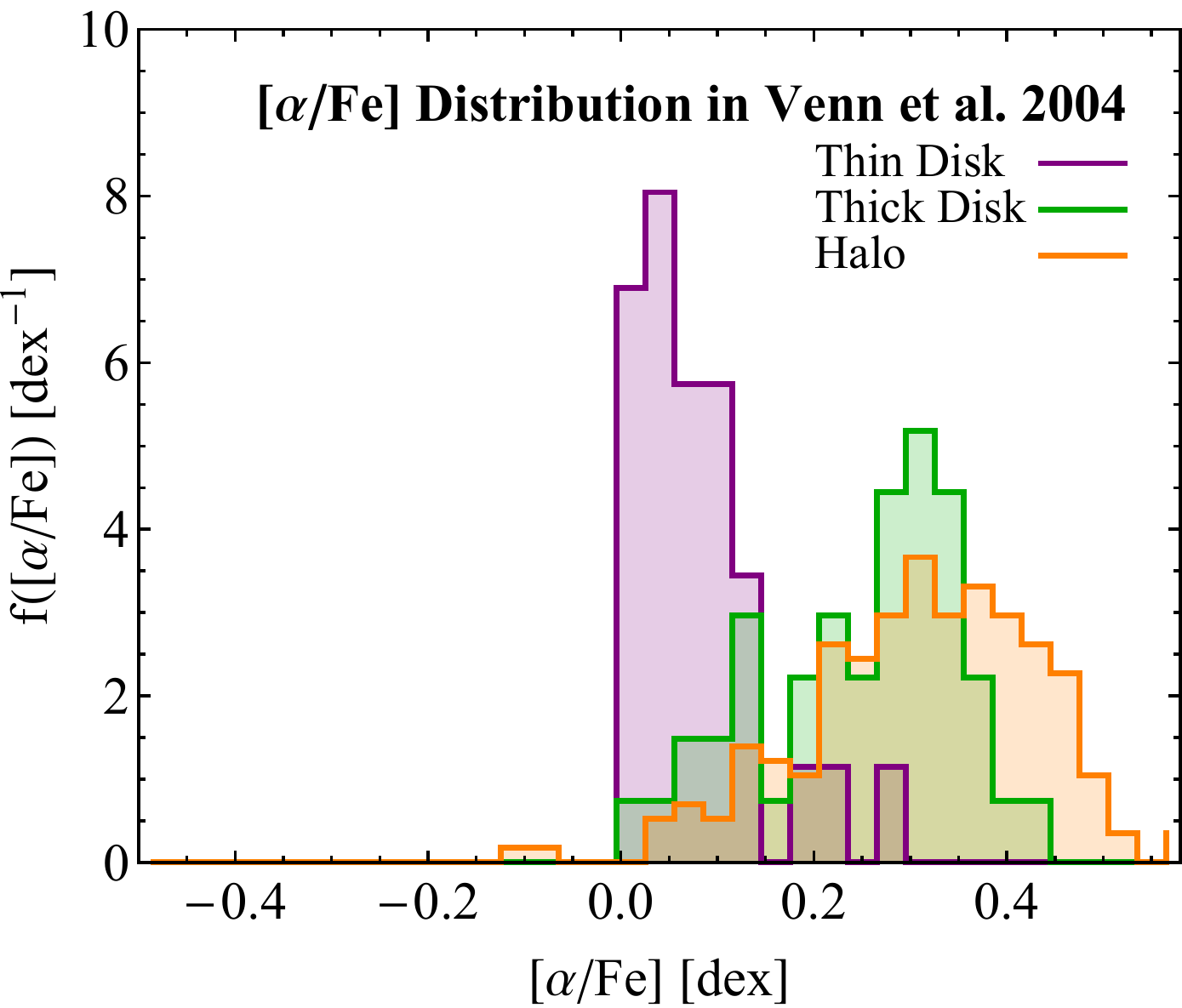}
\end{center}
\caption{
(Left) Iron abundance $\FeH$ distributions and (right) $\alphaFe$ distributions. (Top two rows) \Eris stars within galactocentric radii $6 \leq r \leq 10$~kpc, as a function of vertical distance off the disk plane.  As one moves off the plane, the disk component (centered at \FeH$\sim-0.7$) becomes less prominent, and the low-metallicity tail containing the halo stars becomes more pronounced. Similarly, the $\alphaFe$ distribution increases from $\alphaFe \sim 0.2$ to peak around $\alphaFe \sim 0.5$ for higher values of $z$.  The top row shows the original \textsc{Eris} abundances, while the middle row shows the updated distributions using a more modern IMF and updated yields.  See text for more details. (Bottom row) Chemical properties of stars from \Ref{2004AJ....128.1177V}, divided by the stellar component that they are most likely drawn from: thin disk, thick disk or the stellar halo. 
}
\label{fig:metallicity_dist}
\end{figure}

The relative contribution of Fe and O depends on the assumptions made for the IMF and mass yields.  Let us consider how these values change taking the more modern Chabrier~(2003) IMF~\cite{2003ApJ...586L.133C} and the mass yields of~\Ref{2007PhR...442..269W,Iwamoto:2000as}.  This follows the prescription used for the \textsc{Agora} project~\cite{Kim:2013jpa}.  The Chabrier mass function is
\begin{equation}
\Phi_\text{Chabrier}(m) = M_s \times N_C \times \left\{
        \begin{array}{ll}
           m^{-1} \, e^{ -(\log m - \log m_c)^2/2\sigma^2} & \quad m < 1\, \text{M}_\odot \\
            0.283 \, m^{-2.3} & \quad m \geq 1\, \text{M}_\odot \, ,
        \end{array}
    \right.
\end{equation}
where $m_c = 0.08$~M$_\odot$, $\sigma = 0.69$, and the normalization $N_C = 0.843$ is chosen such that $\int m \, \Phi_\text{Chabrier} \text{d}m = 1$~M$_\odot$ for $0.1\leq m\leq 100$~M$_\odot$.  In this case, 64.7 and 10.4 Type~II and Ia SNe are produced per star particle, respectively.  This assumes that the SN Ia rate is 14\% of the total, in agreement with recent estimates by~\Ref{2012MNRAS.426.3282M}.  The iron and oxygen yields for a Type~II explosion are
\begin{eqnarray}
M_\text{Fe} &=&  0.375 \, e^{-17.94/m} \, \text{M}_\odot \nonumber \\
M_\text{O} &=& 27.66 \, e^{-51.81/m}\, \text{M}_\odot \, ,
\end{eqnarray} 
whereas a Type~Ia SN produces 0.63~M$_\odot$ and 0.14~M$_\odot$ of Fe and O, respectively.  Therefore, 6.74~M$_\odot$ of Fe and 80.2~M$_\odot$ of O are produced by Type~II events, whereas 6.53~M$_\odot$ of Fe and 1.45~M$_\odot$ of O are produced by Type~Ia events.  These assumptions yield 3.10 times more Fe and 
2.81 times more O (per star particle) than the \textsc{Eris} values.  For the metallicities and alpha abundances presented in the Letter, we use the more modern relations, which shift the \textsc{Eris} metallicities by the following amounts:
\begin{equation}
\Delta \left(\FeH\right) = + 0.49 \quad \text{and} \quad \Delta \left(\alphaFe\right) = -0.044 \, . 
\end{equation}
The updated abundance distributions are shown in the middle panel of Fig.~\ref{fig:metallicity_dist} and can be compared to the original distributions in the top panel.  The updated abundances are used in all results presented in this Letter.  As a reference, we also show the observed $\FeH$ and $\alphaFe$ distributions from~\Ref{2004AJ....128.1177V} in the bottom panel.  We include the stars that have a $90\%$ probability of belonging to either the thin disk, thick disk or the stellar halo.

\end{document}